\documentstyle[12pt,psfig]{article}

\begin{document}

\begin{titlepage}
\begin{center}

{\Large\bf{Searching Saturation in $eA$ Processes}}
\\[5.0ex]
{ \Large \it{ V. P.  Gon\c{c}alves $^{**}$\footnotetext{$^{**}$E-mail:barros@ufpel.tche.br} 
}} \\[1.5ex]
{\it Instituto de F\'{\i}sica e Matem\'atica, Univ. Federal de Pelotas}\\
{\it Caixa Postal 354, 96010-900, Pelotas, RS, BRAZIL}\\[5.0ex]
\end{center}

{\large \bf Abstract:}

The high density effects should be manifest at small $x$ and/or large nuclei. In this letter we consider the behavior of  nuclear structure function $F_2^A$ slope in the kinematic region which could be explored in the future  $eA$  colliders as a search of these effects. 
We verify that the high density  implies that the maximum value of the slope occurs at large values of the photon virtuality, {\it i. e.} in a perturbative regime, and is dependent of the number of nucleons $A$ and  energy. Our conclusion is that the measurement of this observable will allow to explicit the saturation.

\vspace{1.5cm}



\end{titlepage}

\label{int}

The physics of high-density QCD (hdQCD) has become an increasingly active
subject of research, both from experimental and theoretical points of view.
Presently, and in the near future, the collider facilities such as the DESY 
collider HERA ($ep$, $eA$), Fermilab Tevatron ($p\overline{p}$, $pA$), BNL
Relativistic Heavy Ion Collider (RHIC) ($eA$, $AA$), and CERN Large Hadron
Collider (LHC) ($p\overline{p}$, $AA$) will be able to probe new regimes of
dense quark gluon matter at very small Bjorken $x$ or/and at large $A$, with
rather different dynamical properties. In these experiments, be it because
of high energies in $ep$ collisions or because of intrinsically higher
numbers of partons in $eA$ and $AA$ collisions, QCD effects are dominated by
the large number of gluons involved. The description of these processes is
directly associated with a correct gluonic dynamics in this kinematical
region.

Theoretically, at small $x$ and/or large $A$ we expect the transition of the
regime described by the linear dynamics (DGLAP, BFKL) (For a review, see
e.g. Ref. \cite{cooper}), where only the parton emissions are considered,
for a new regime where the physical process of recombination of partons
become important in the parton cascade and the evolution is given by a
nonlinear evolution equation. This regime is characterized by the limitation
on the maximum phase-space parton density that can be reached in the hadron
wavefunction (parton saturation) and very high values of the QCD field
strength $F_{\mu \nu} \approx 1/g$ \cite{mue}. In this case, the number of
gluons per unit phase space volume practically saturates and at large
densities grows only very slowly (logarithmically) as a function of the
energy. At this moment, there are many approaches in the literature that
propose distinct evolution equations for the description of the gluon
distribution in high density limit \cite{qiu,ayala1} \cite{jamal,kov}. In
general these evolution equations resum powers of the function $\kappa
(x,Q^2) \equiv \frac{3 \pi^2 \alpha_s A }{2 Q^2 } \frac{ xg(x,Q^2)}{\pi R^2_A%
} $, which represents the probability of gluon-gluon interaction inside the
parton cascade, matching

\begin{itemize}
\item  the DLA limit of the DGLAP evolution equation in the limit of low
parton densities $(\kappa \rightarrow 0)$;

\item  the GLR equation and the Glauber-Mueller formula as first terms of
the high density effects.
\end{itemize}

The main differences between these approaches occurs in the limit of very
large densities, where all powers of $\kappa $ should be resumed. Although
the complete demonstration of the equivalence between these formulations in
the region of large $\kappa $ is still an open question, some steps in this
direction were given recently \cite{npbvic,kovner}.

Considering that the condition $\kappa = 1$ specifies the critical line,
which separates between the linear regime $\kappa \ll 1$ and the high
density regime $\kappa \gg 1$, we can define the saturation momentum scale $%
Q_s$ given by 
\begin{eqnarray}
\kappa = 1 \Rightarrow Q_s^2 (x; A) = \frac{3 \pi^2 \alpha_s A }{2 } \frac{
xg(x,Q_s^2(x;A))}{\pi R^2_A}\,\,\,,  \label{sat}
\end{eqnarray}
below which the gluon densities reach their maximum value (saturates). At any
value of $x$ there is a value of $Q^2 = Q_s^2(x)$ in which the gluonic
density reaches a sufficiently high value that the number of partons stops
the growth. This scale depends on the energy of the process and the atomic
number of the colliding nuclei, determining  the typical intrinsic momenta
associated with quanta in the nuclear wavefuntion. These quanta go on shell
in a collision, and eventually produce a large multiplicity of particles. At
very high energies, the saturation scale is the only scale in the problem,
and one can estimate that the typical momenta of the particles produced in
the collision is at this scale. If this momenta is large enough, one can
approach the saturation regime using perturbation theory. Recently, the high
density effects in $AA$ collisions were considered \cite{venu}, verifying
that the equilibration time, the initial temperature and the chemical
potential have a strong functional dependence on the initial gluon
saturation scale $Q_s$.

The deeply inelastic scattering experiments at HERA revealed that structure
functions grows rapidly at small $x$ and large $Q^2$. Therefore, at some
value of $x$, for a fixed $Q^2$, it is expected that the parton distribution
will saturate, leading to a weaker growth of the structure function.
However, although these high density effects should be present in the $ep$
HERA kinematical region, the current limited $x$ range available at HERA
makes it difficult to distinguish between the predictions of the linear and
nonlinear dynamics. Basically, the same data compatible with the high
density approaches can be described from a different point of view without
the nonlinear effects in the standard DGLAP evolution equation for $%
Q^2>1\,GeV^2$ and the soft phenomenology for $Q^2<1\,GeV^2$ \cite{f2h1ze}.
In Ref. \cite{golec} the authors has analyzed the position of the critical
line, verifying that going along the critical line from $x=10^{-4}$ to $%
x=10^{-5}$ the saturation scale increases from approximately $1\,GeV^2$ up to 
$2\,GeV^2$, {\it i. e.} the saturation scale is approximately constant in
the $ep$ HERA kinematical region, allowing that the high density effects can
be absorbed to a large  extent in the initial conditions of the linear dynamics 
\cite{mrst,grv98}.

Here we analyze the high density effects in $eA$ processes, where we can
study the dynamics of QCD at high densities and at zero temperature, raising
questions complementary to those addressed in the search for a quark-gluon
plasma in high-energy heavy ion collisions. The nucleus in this process
serves as an amplifier for nonlinear phenomena expected in QCD at small $x$,
obtaining at the assessable energies at HERA and RHIC with an $eA$ collider
the parton densities which would be probed only at energies comparable to
LHC energies with an $ep$ collider. Our goal is to address the boundary
region between the linear and nonlinear dynamics and identify possible
signatures in the behavior of the nuclear structure function slope. As the
behavior of this quantity is strongly dependent of the gluon distribution,
we expect distinct behaviors at $Q^2>Q_s^2$ and $Q^2<Q_s^2$ with the
transition point dependent of the values of $x$ and $A$. We assume that in
the boundary region, the parton density is sufficiently large to invalid the
descriptions which use the linear dynamics, but small to consider a general
approach which resum all powers of $\kappa $. Basically, we will consider
the high density effects in the kinematical region where the predictions
using the Glauber-Mueller approach \cite{ayala1} are a good approximation,
and our analysis in principle is not model dependent. This study is
motivated by our results in Ref. \cite{df2a}, where we show that the deep
inelastic scattering on nuclear targets is a very good place to look for
saturation if we consider the behavior of the nuclear structure function
slope, and demonstrate the distinct predictions from DGLAP and high density
approaches. Here we extend the previous analyses for the energies of $eA$
processes at RHIC and at HERA, as well as for the possible colliding nuclei in
these experiments, analyzing the $x$ and $A$ dependence of the saturation
scale (For a related discussion see Ref. \cite{mcl}).

The deep inelastic scattering $eA\rightarrow e+X$ is characterized by a
large electron energy loss $\nu $ (in the target rest frame) and an
invariant momentum transfer $q^2\equiv -Q^2$ between the incoming and
outgoing electron such that $x=Q^2/2m_N\nu $ is fixed. It is usually
interpreted in a frame where the nucleus is going very fast. In this case
the high density effects are the result of an overlap in the longitudinal
direction of the parton clouds originated from different bound nucleons \cite
{qiu}. It corresponds to the fact that small $x$ partons cannot be localized
longitudinally to better than the size of the nucleus. Thus low $x$ partons
from different nucleons overlap spatially creating much larger parton
densities than in the free nucleon case. This leads to a large amplification
of the nonlinear effects expected in QCD at small $x$. In the target rest
frame, the electron-nucleus scattering can be visualized in terms of the
propagation of a small $q\overline{q}$ pair in high density gluon fields
through much larger distances than it is possible with free nucleons. In
terms of Fock states we then view the $eA$ scattering as follows: the
electron emits a photon ($|e \rangle \rightarrow |e\gamma \rangle $) with $E_\gamma =\nu $
and $p_{t\,\gamma }^2\approx Q^2$, after the photon splits into a $q%
\overline{q}$ ($|e\gamma \rangle \rightarrow |eq\overline{q} \rangle $) and typically
travels a distance $l_c\approx 1/m_Nx$, referred as the 'coherence length',
before interacting in the nucleus. For small $x$, the photon converts to a
quark pair at a large distance before it interacts with the target.
Consequently, the space-time picture of the DIS in the target rest frame can
be viewed as the decay of the virtual photon at high energy (small $x$) into
a quark-antiquark pair long before the interaction with the target. The $q%
\overline{q}$ pair subsequently interacts with the target. It allows to
factorize the total cross section between the wavefunction of the photon
and the interaction cross section of the quark-antiquark pair with the
target. The photon wavefunction is calculable and the interaction cross
section is modelled. Moreover, the cross sections for transverse and longitudinal
photons are most conveniently written in a mixed representation. The two
transverse directions are treated in the coordinate space, while the
longitudinal direction is described in the momentum representation. Let $%
\vec{r}_t$ be the two dimensional vector pointing from the quark to the
antiquark in the transverse plane and $z$ the fraction of photon energy $\nu 
$ carried by the quark. The momentum fraction of the antiquark is then $1-z$%
. The nuclear structure function reads \cite{nik} 
\begin{eqnarray}
F_2^A(x,Q^2)=\frac{Q^2}{4\pi \alpha _{em}}\int dz\int \frac{d^2\vec{r}_t}\pi
|\Psi (z,\vec{r}_t)|^2\,\sigma ^{q\overline{q}+A}(z,\vec{r}_t)\,\,,
\label{f2a}
\end{eqnarray}
where $\Psi (z,\vec{r}_t)$ is the light-cone wavefunction for the transition 
$\gamma ^{*}\rightarrow q\overline{q}$ and we have assumed the dominance of
the transverse photon polarization. The cross section for scattering a $q%
\overline{q}$ - dipole off the nucleus is denoted by $\sigma ^{q\overline{q}%
+A}(z,\vec{r}_t)$. In the pure perturbative regime the reaction is mediated
by single gluon exchange which changes into multi-gluon exchange when the
saturation region is approached.

We estimate the high density effects considering the Glauber multiple
scattering theory \cite{chou}, which was probed in QCD \cite{muegla}. The
nuclear collision is analyzed as a succession of independent collisions of
the probe with individual nucleons within the nucleus, which implies that 
\begin{eqnarray}
F_2^A(x,Q^2)=\frac{Q^2}{4\pi \alpha _{em}}\int dz\int \frac{d^2\vec{r}_t}\pi
|\Psi (z,\vec{r}_t)|^2\,\int \frac{d^2\vec{b}_t}\pi \,2\,[1-e^{-\sigma ^{q%
\overline{q}+N}(z,\vec{r}_t)S(\vec{b}_t)}]\,\,,  \label{siga}
\end{eqnarray}
where $\vec{b}_t$ is the impact parameter, $S(\vec{b}_t)$ is the profile
function and $\sigma ^{q\overline{q}+N}$ is the dipole cross section off the
nucleons inside the nucleus, which is proportional to the pair separation
squared $r_t^2$ and the nucleon gluon distribution $xg(x,1/r_t^2)$. The
expression (\ref{siga}) represents is the Glauber-Mueller formula for the
nuclear structure function (see \cite{ayala1} for details).
The main characteristic of the Glauber-Mueller formula
is that for a large $Q^2$ it reduces to the standard small $x$ DGLAP
expression, while at small $Q^2$ it goes to zero as $Q^2logQ^2$, predicting
a transition in the behavior of $F_2^A$ at an intermediate value of $Q^2$. A similar
expression was used in Ref. \cite{golec} for a phenomenological analysis of
the $ep$ process, disregarding the geometrical structure of the collision,
the $\vec{b}_t$ dependence, and assuming $xg\propto x^{-\lambda
}\,\,(\lambda >0)$, resulting a very good description of the HERA data.

Using a gaussian profile function, we can derive the slope of the nuclear
structure function directly from the expression (\ref{siga}), resulting \cite
{df2a} 
\begin{eqnarray}
\frac{d F_2^A(x,Q^2)}{dlog Q^2} = \frac{R_A^2 Q^2}{2\pi^2} \sum_1^{n_f}
\epsilon_i^2 \{C + ln(\kappa_q(x, Q^2)) + E_1(\kappa_q(x,Q^2))\}\,\,,
\label{diseik2}
\end{eqnarray}
where $\kappa_q = (2 \alpha_s\,A/3R_A^2)\,\pi\,r_t^2\,  xg(x,\frac{1}{r_t^2})
$, $A$ is the number of nucleons, $R_A^2$ is the mean nuclear radius, $C$ is
the Euler constant and $E_1$ is the exponential function (see \cite{df2a}
for details). The expression (\ref{diseik2})  predicts the $x, \,Q^2$ and $A$
dependence of the high density effects for the $F_2^A$ slope. Similarly to
the $F_2^A$ we expect that a turnover, in the behavior of the $F_2^A$ slope, should
occur for the saturation scale.

Here we estimate the high density effects for the $F_2^A$ slope in the
kinematic regions which could be explored in $eA$ colliders at HERA and
RHIC, as well as for some typical values of the number of nucleons $A$. We
use as input in our calculations the GRV95 parameterization \cite{grv95} for
the nucleon gluon distribution, since we expect that the nonlinear
corrections to the DGLAP evolution equation probably seen at HERA are not
obscured by it (see Ref. \cite{raio} for a full analysis).

As discussed in Ref. \cite{hera96}, in a first moment the experiments should
be carried with nuclear targets with the ratio $Z/A$ equal to $1/2$. Hence
the energy for  each nucleon in a deep inelastic collision will be half that
in an $ep$ collision. For instance, for $eA$ processes at HERA, we will have
that the energy per nucleon will be about $410\,GeV$. Assuming the current
value of $27.6\,GeV$ for electron energy results $W\approx 110\,GeV$ at HERA
and $W\approx 75$ at RHIC, where energy/nucleon $=200\,GeV$. Therefore, we
will consider the possibility of  electron scattering with  carbon ($A=12
$) and  sulfur ($A=32$) targets, which satisfy the above condition, and
extrapolate our analysis for the possibility of collisions with $Au$ targets
($A=197$).

Following \cite{golec}, we consider that the Bjorken variable $x$ and the
photon virtuality $Q^2$ are related by the expression $x=Q^2/W^2$, where $W$
is the $\gamma ^{*}p$ c.m. energy, and calculate the $x$ and $Q^2$
dependences for $W$ and $A$ fixed. This approach is inspired by the recent
analysis carried by ZEUS \cite{df2zeus} to stipulate the deviation from the
conventional perturbative QCD framework at low values of $Q^2$. The Figs. 
\ref{fig1} and \ref{fig2} shows the logarithmic $Q^2$ slope of $F_2^A$
plotted for fixed $W^2$ and different $A$ as a function of $Q^2$ and $x$,
respectively. The remarkable property of the plots is the presence of a distinct maximum for
each slope, dependent of the energy and the number of nucleons considered.
We can see that at fixed $A$, if we increase the value of the energy, the
maximum value of the slope occur at larger values of $Q^2$ and smaller
values of $x$. A similar behavior for the proton structure function slope
was obtained in Ref. \cite{golec}. The remarkable properties of the
collisions with nuclei targets are the large values of $Q^2$ for the
turnover of $F_2^A$ slope and the large shift of the turnover at large
values of $Q^2$ with the growth of the energy. In Table \ref{tab1} we
present explicitly the $A$ and $W$ dependences of the turnover in $F_2^A$
slope. We verify that, at fixed $W$, the turnover is displaced at larger
values of $x$ and $Q^2$ if we increase the number of nucleons $A$. These
behaviors can be understood intuitively. The turnover is associated with the
regime in which the partons in the nucleus form a dense system with mutual
interactions and recombinations, with a transition between the linear and
nonlinear regime at the saturation scale [Eq. (\ref{sat})]. As the partonic
density growth at larger values of the number of nucleons $A$ and smaller
values of $x$ we have that, at fixed $A$, the saturation scale $Q_s^2$ will
increase at small values of $x$, since this is directly proportional to the
gluon distribution. Moreover, at fixed energy $W$, the same density at $%
A=12,32,197$ will be obtained at larger values of $x$ and $Q^2$. These
properties of the high density effects are verified in the $F_2^A$ slope.
The main result of our analysis is that the saturation should occur already
at rather small distances (large $Q_s^2$) well below where soft dynamics is
supposed to set in, justifying the use of perturbative QCD to approach a
highly dense system.

Our main conclusion is that deeply inelastic scattering of electrons off
nuclei at high energies can determine whether parton distributions saturate.
The analysis  of the nuclear structure function slope at $eA$ HERA
and/or RHIC energies will allow to establish the presence of a high density
system and the behavior of the saturation scale. Moreover, we predict that
the transition between the linear and nonlinear regimes in $eA$ processes at
high energies will occur in a perturbative regime, justifying the current
pQCD approaches. We hope that our estimates will be useful for planning of
future $eA$ experiments.

\section*{Acknowledgments}

This work was partially financed by CNPq, BRAZIL.

\newpage 

\section*{Tables}

\vspace{2.0cm}

\begin{table}[h]
\begin{center}
\begin{tabular}{||l||l|l||l|l||l|l||}
\hline\hline
& W = 50 GeV &  & W = 75 GeV &  & W = 115 GeV &  \\ 
& $x$ & $Q^2 $ & $x$ & $Q^2 $ & $x$ & $Q^2 $ \\ \hline\hline
$A = 12$ & $0.620 \times 10^{-3}$ & $1.55$ & $0.382 \times 10^{-3}$ & 2.15 & 
$0.231 \times 10^{-3}$ & 3.05 \\ \hline
$A = 32$ & $0.740 \times 10^{-3}$ & $1.85$ & $0.462 \times 10^{-3}$ & 2.6 & $%
0.287 \times 10^{-3}$ & 3.8 \\ \hline
$A = 197$ & $0.11 \times 10^{-2}$ & $2.75$ & $0.649 \times 10^{-3}$ & 3.65 & 
$0.389 \times 10^{-3}$ & 5.15 \\ \hline\hline
\end{tabular}
\end{center}
\caption{The $A$ and $W$ dependences of the turnover predicted in the $F_2^A$
slope.}
\label{tab1}
\end{table}

\newpage

\section*{Figure Captions}

\begin{figure}[tbp]
\centerline{\psfig{file=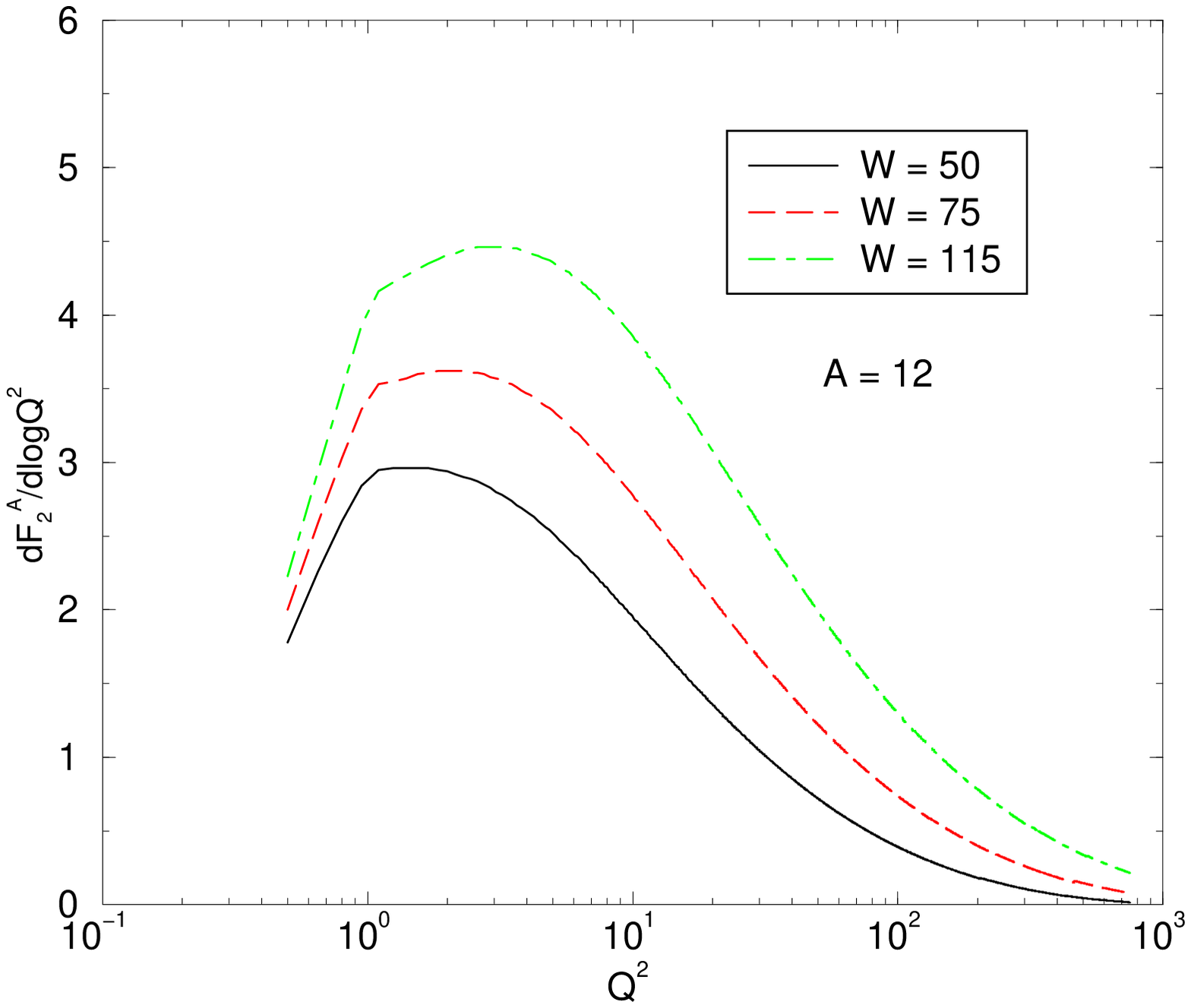,width=70mm}} \centerline{%
\psfig{file=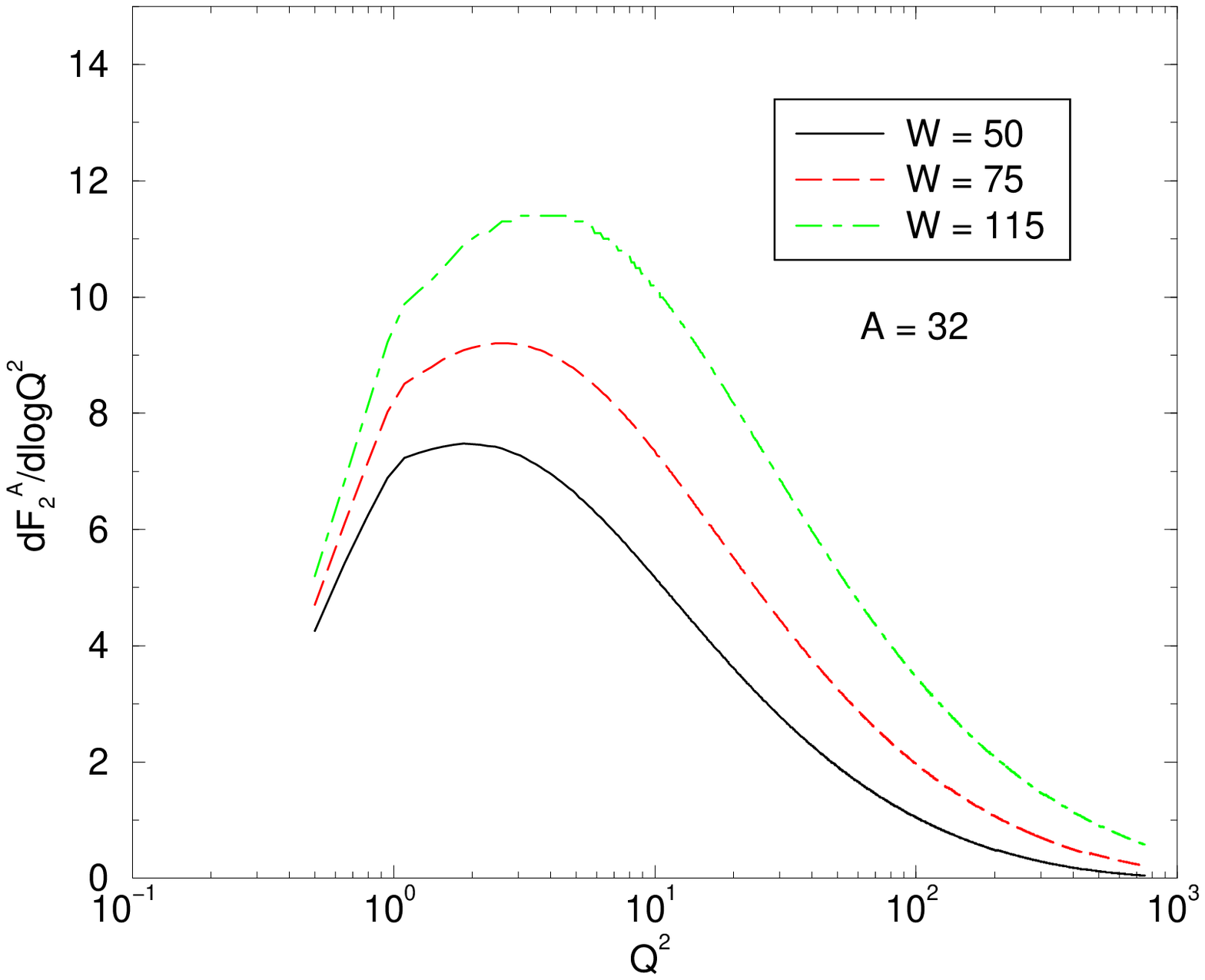,width=70mm}} \centerline{%
\psfig{file=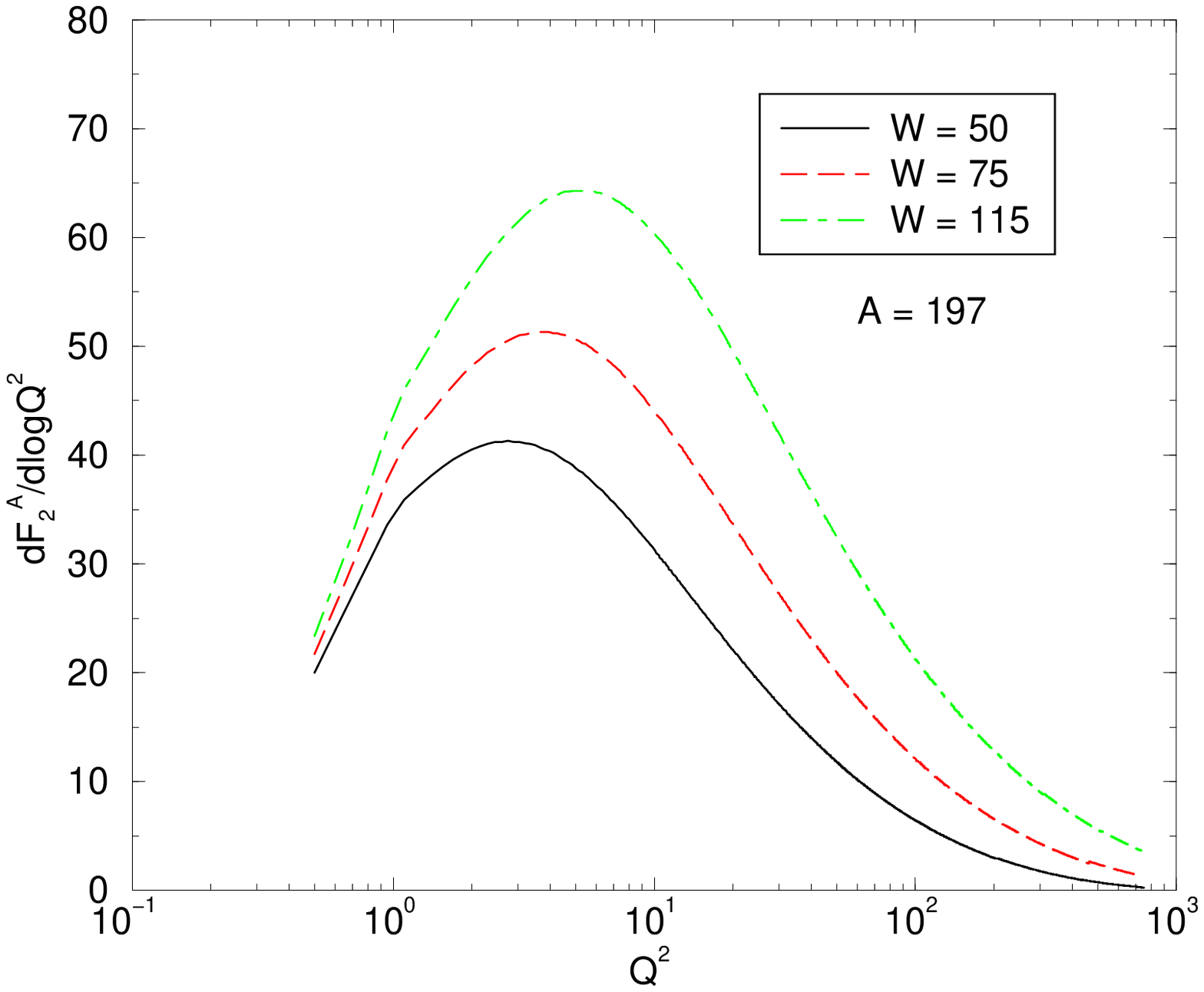,width=70mm}} 

\caption{ }
\label{fig1}
\end{figure}

\vspace{1.0cm} Fig. \ref{fig1}: The logarithmic $Q^2$ slope of $F_2^A$ as
function of the variable $Q^2$ at different values of $A$ and $W$. See text.

\begin{figure}[tbp]
\centerline{\psfig{file=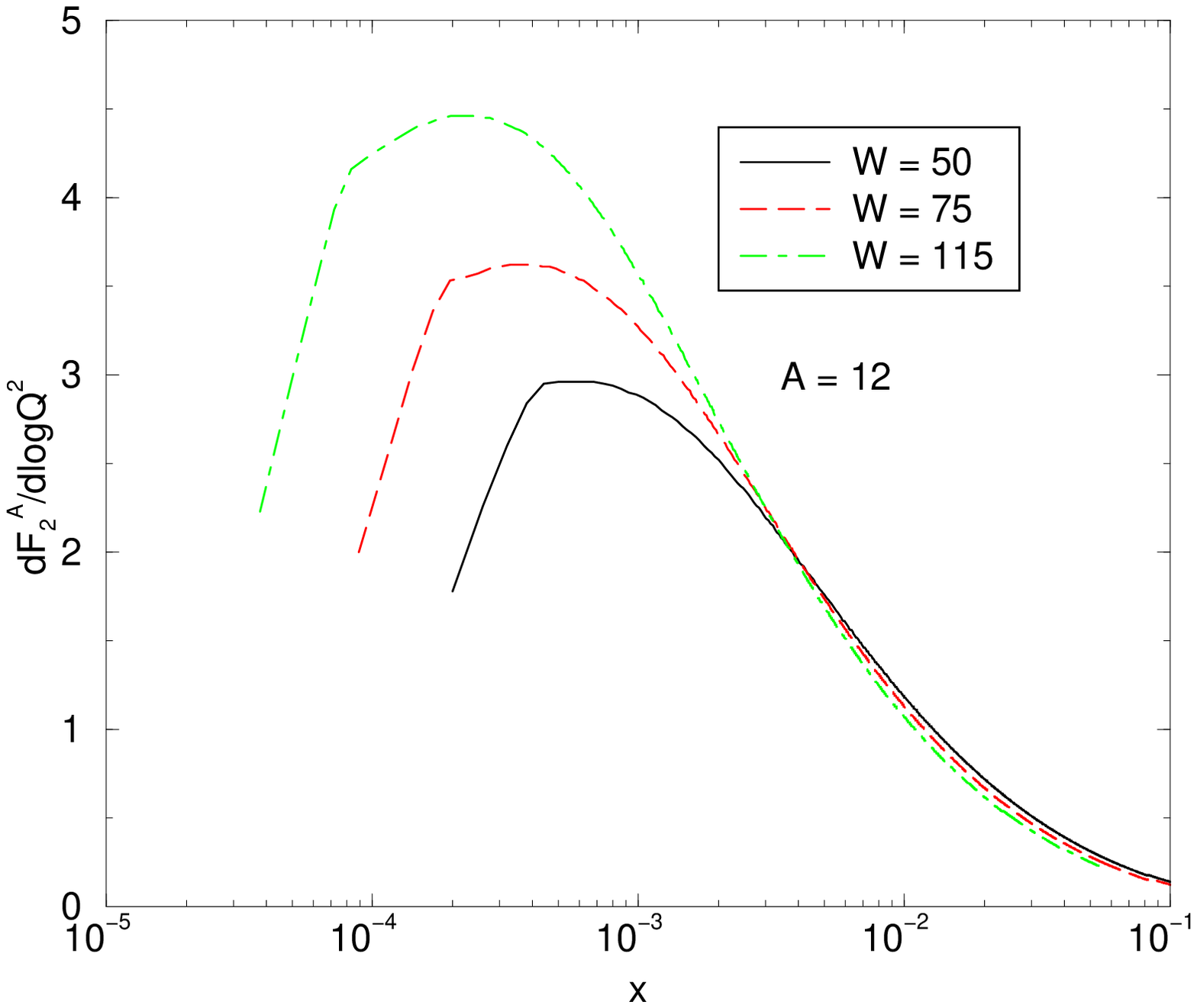,width=70mm}} \centerline{%
\psfig{file=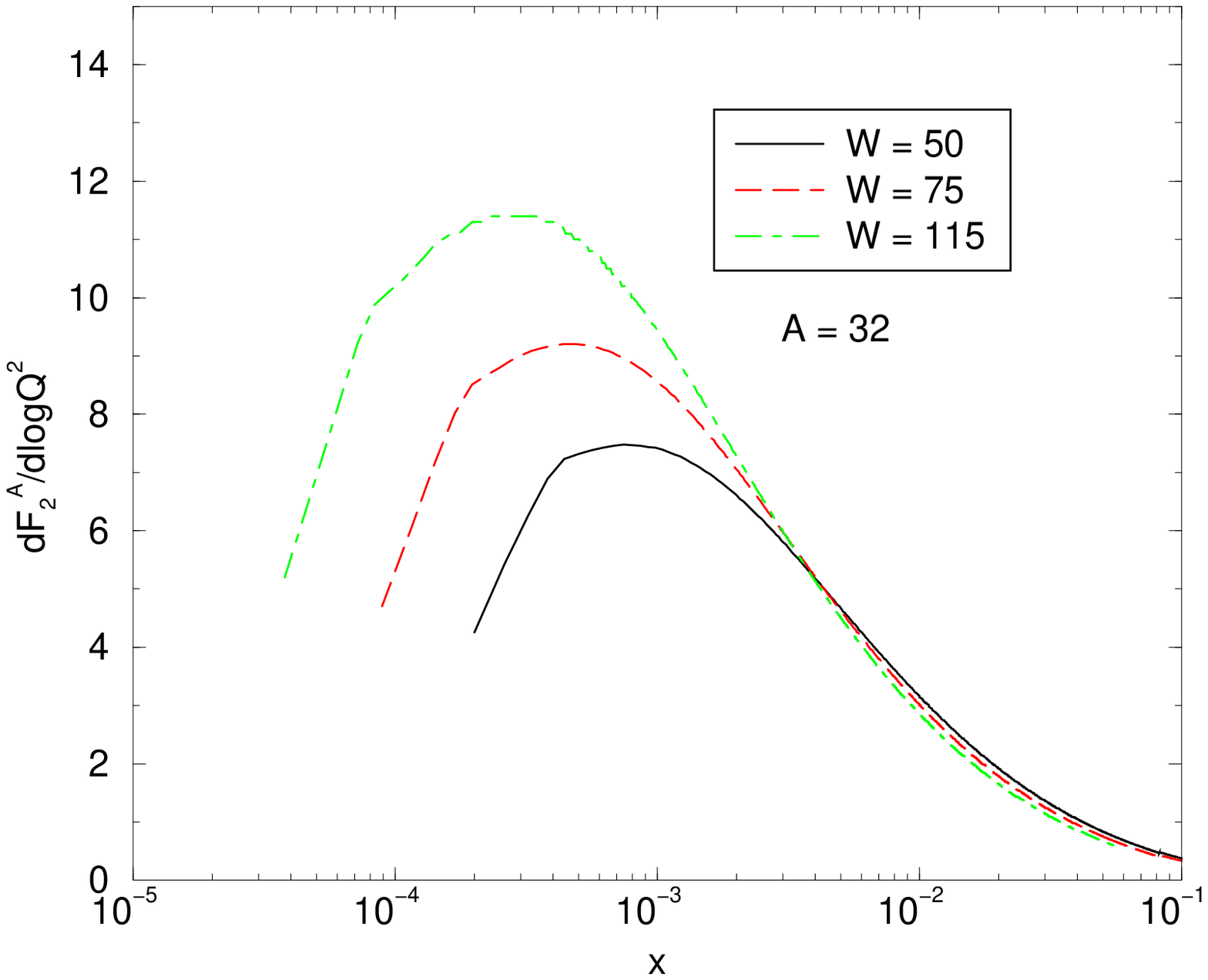,width=70mm}} \centerline{%
\psfig{file=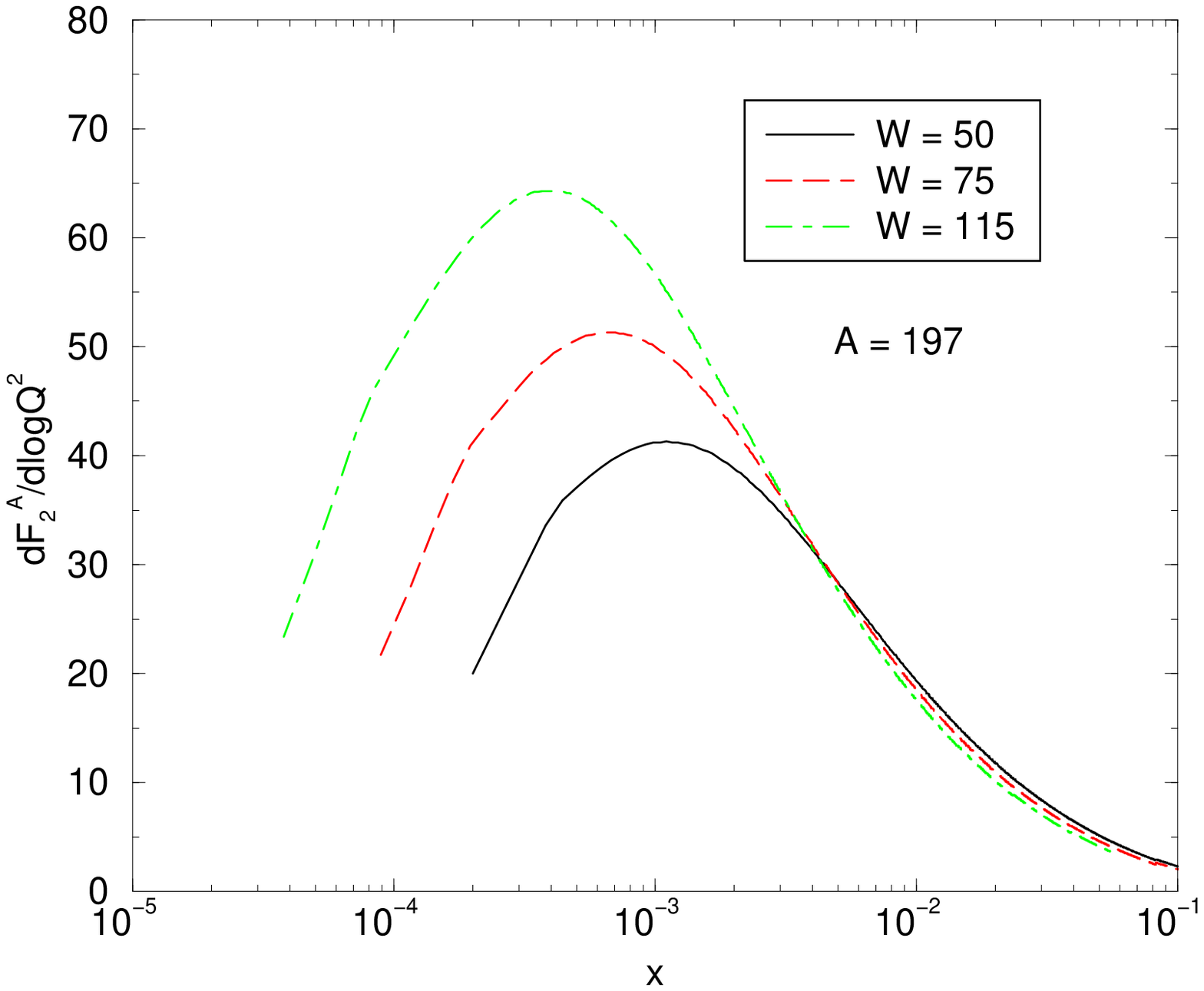,width=70mm}} 

\caption{ }
\label{fig2}
\end{figure}

\vspace{1.0cm} Fig. \ref{fig2}: The logarithmic $Q^2$ slope of $F_2^A$ as
function of the variable $x$ at different values of $A$ and $W$. See text.

\newpage

\end{document}